\documentclass[fleqn,10pt]{wlscirep}
\usepackage[utf8]{inputenc}
\usepackage[T1]{fontenc}
\title{Upper critical field and superconductor-metal transition in ultrathin niobium films}

\author[1*]{Iryna Zaytseva}
\author[1]{Aleksander Abaloszew}
\author[1]{Bruno C. Camargo}
\author[1]{Yevgen Syryanyy}
\author[1*]{Marta Z. Cieplak}
\affil[1]{Institute of Physics, Polish Academy of Sciences, Aleja Lotnik\'{o}w 32/46, PL-02668 Warsaw, Poland}

\affil[*]{zaytseva@ifpan.edu.pl}
\affil[*]{marta@ifpan.edu.pl}

\begin{abstract}

Recent studies suggests that in disordered ultrathin films superconducting (SC) state may be intrinsically inhomogeneous. Here we investigate the nature of SC state in ultrathin Nb films, of thickness $d$ ranging from 1.2 nm to 20 nm, which undergo a transition from amorphous to polycrystalline structure at the thickness $d \simeq 3.3$ nm. We show that the properties of SC state are very different in polycrystalline and amorphous films. The upper critical field ($H_{c2}$) is orbitally limited in the first case, and paramagnetically limited in the latter. The magnetic field induced superconductor-metal transition is observed, with the critical field approximately constant or decreasing as a power-law with the film conductance in polycrystalline or amorphous films, respectively. The scaling analysis indicates distinct scaling exponents in these two types of films. Negative contribution of the SC fluctuations to conductivity exists above $H_{c2}$, particularly pronounced in amorphous films, signaling the presence of fluctuating Cooper pairs. These observations suggest the development of local inhomogeneities in the amorphous films, in the form of proximity-coupled SC islands. An usual evolution of SC correlations on cooling is observed in amorphous films, likely related to the effect of quantum fluctuations on the proximity-induced phase coherence.

\end{abstract}
\begin{document}

\flushbottom
\maketitle

\section*{Introduction}

The SC state in homogeneously disordered, two dimensional (2D) films may be destroyed by tuning of various parameters, such as, for example, external magnetic field ($B$), disorder, or doping. The increase of the field induces two, distinctly different transitions: superconductor-insulator transition (SIT) or superconductor-metal transition (SMT) in strongly or weakly disordered films, respectively \cite{Gantmakher2010}. The nature of these transitions is not fully understood. Two early scenarios have been considered, fermionic \cite{Finkel1987} and bosonic \cite{Fisher1990}, predicting that Cooper pairs are either broken or survive the transition, respectively. Fluctuating Cooper pairs (FCP) surviving across the SIT/SMT phase boundary, well into the normal state, may be responsible for many unusual transport phenomena, such as, for example, large positive magnetoresistance (MR) peak, large Nernst effect, negative MR at high magnetic fields, or magnetic field induced reentrant behaviors  \cite{Okuma1998,Samba2004,Steiner2005,Xu2000,Pourret2006,Gantmakher2003,Parker2006,Liu2019}. Some of these phenomena are well described by theories of the FCP, developed over last two decades \cite{Galitski2001,Glatz2011,Varlamov2018}. Another problem considered recently is the development of microscopic inhomogeneities in the form of intertwined SC and non-SC islands near the transition \cite{Feigelman2001,Spivak2001,Dubi2007,Spivak2008,Kapitulnik2019}. The SC islands are indeed observed by scanning tunneling spectroscopy near the SIT \cite{Sacepe2008,Chand2012,Sherman2012,Kamlapure2013}, while conflicting tunneling results are reported for the SMT \cite{Shabo2016,Ganguly2017}. Such islands influence the behavior of SC correlations near the transition, as suggested by studies of superconductivity in arrays of SC islands placed on metallic films \cite{Eley2012,Han2014}.

Here, we evaluate the $B$-induced SMT (with $B$ perpendicular to the film plane) in the thin niobium (Nb) films. This issue has not been investigated before, despite many studies devoted in the past to transport properties of Nb films
\cite{Mayadas1972,Gershenzon1983,Park1986,Dalrymple1986,Quateman1986,Gurvitch1986,Hikita1990,Hsu1992,Yoshii1995,Delacour2011}.
The films, with thickness $d$ between 1.2 nm and 20 nm, are sandwiched between two Si barrier layers of 10 nm to prevent oxidation, as described recently \cite{Zaytseva2014}. With decreasing $d$ the films undergo the change of structure from polycrystalline to amorphous at $d \simeq 3.3$ nm, accompanied by the growing contribution of electron carriers to the conduction, which in the bulk Nb is dominated by holes. X-ray Photoelectron Spectroscopy evaluation \cite{Demchenko2017} suggests that this effect is due to strong surface scattering of holes, enhanced by a small admixture of Si ions (at the level of 5-10 at.\%) into the Nb layer closest to the interface.

The most important finding in the present study is the strong indication of the development of proximity-coupled SC islands in the amorphous films, what results in SC state with properties distinctly different from those observed in polycrystalline films. These distinct properties include upper critical field ($H_{c2}$), the scaling properties in the vicinity of the SMT, and the negative contribution of SC fluctuations to conductivity above $H_{c2}$, particularly large in amorphous films, consistent with the theories of the FCP \cite{Galitski2001,Glatz2011,Varlamov2018}. An unusual evolution of the SC correlations is observed in the amorphous films on lowering of temperature, which we propose to explain by the influence of quantum fluctuations on the proximity-induced phase coherence.

\section*{Results}

\subsection*{Magnetoresistance}

The details on the film preparation and calibration of $d$ have been described previously \cite{Zaytseva2014} (a brief summary is given in Supplementary information). Since the dependence of the resistance per square, $R_{sq}$, on $d$ is mostly monotonic, we use $d$ values to label different films. However, some fluctuations of film thickness ($\sim$ 10\%) are unavoidable for small $d$, contributing to scattering of $R_{sq}$ values; this is illustrated by two different films with $d = 1.3$ nm, labeled "a" and "b". Table \ref{tab} lists all films in the set, together with parameters determined in this study.

\begin{table}[h]
 \begin{centering}
  \begin{tabular}{|c|c|c|c|c|c|c|c|}
  \hline \begin{tabular}{c}
          $d$ \\
          (nm) \\
         \end{tabular}
&\begin{tabular}{c}
    $R_{N}$ \\
    ($\Omega$) \\
   \end{tabular}
&\begin{tabular}{c}
  $T_c$ \\
  (K)\\
 \end{tabular}
&\begin{tabular}{c}
  $T_{c}^{on}$ \\
  (K)\\
 \end{tabular}
&\begin{tabular}{c}
    $l$ \\
    (nm) \\
   \end{tabular}
&\begin{tabular}{c}
    $B_c$ \\
    (T) \\
   \end{tabular}
&\begin{tabular}{c}
 $\mu_0 H_{c2}(0)$ \\
 (T)\\
 \end{tabular}
&\begin{tabular}{c}
    $\xi(0)$ \\
    (nm) \\
   \end{tabular} \\

  \hline 1.2  & 2712  & 0    & 0.35 & 0.15 & 0.29 &      &     \\
  \hline 1.3b & 1873  & $\sim 0.8$   & $\sim 1.3$ & 0.21 & 1.19 &      &     \\
  \hline 1.3a & 1503  & 0.86 & 1.5 & 0.26 & 1.45 & 1.46 & 15.0\\
  \hline 1.4  & 1653  & 1.19 & 1.78 & 0.22 & 1.46 & 1.5  & 14.8\\
  \hline 2.2  & 886   & 1.63 & 2.28 & 0.26 & 2.52 & 2.41 & 11.7\\
  \hline 3.2  & 578.8 & 2.38 & 3.13 & 0.27 & 3.48 & 3.28 & 10.0\\
  \hline 3.3  & 447   & 2.29 & 2.99 & 0.34 & 3.5  &      &     \\
  \hline 3.9  & 359.3 & 2.67 & 3.39 & 0.36 & 3.4  &      &     \\
  \hline 5.3  & 137.2 & 4.14 & 4.62 & 0.69 & 3.43 & 3.13 & 10.3\\
  \hline 6.7  & 120.9 & 4.75 & 5.22 & 0.62 & 3.5  &      &     \\
  \hline 7.6  & 78.5  & 5.43 & 5.75 & 0.84 &      &      &     \\
  \hline 9.5  & 38.0  & 6.31 & 6.48 & 1.33 & 3.32 & 3.15 & 10.2\\
  \hline 11.3 & 30.8  & 6.16 & 6.29 & 1.37 & 3.3  &      &     \\
  \hline 16   & 6.97  & 7.49 & 7.6 & 3.22 &  -   &      &     \\
  \hline 20   & 7.83  & 7.6  & 7.7 & 2.40 &  -   & 2.64 & 11.2\\
  \hline
  \end{tabular}
  \caption{Parameters of Nb films: thickness $d$, sheet resistance $R_{N}$ at $T = 10$ K, SC transition temperature: $T_c$ (midpoint) and $T_{c}^{on}$ (onset, at $R_{sq}/R_N = 0.95$), mean free path $l$, SMT critical field $B_c$ (at the lowest $T$), upper critical field $\mu_0 H_{c2}(0)$, coherence length $\xi(0)$.}
  \label{tab}
 \end{centering}
\end{table}

First four parameters are: the normal-state resistance measured at $T = 10$ K ($R_N$), the midpoint of the SC transition temperature at $B = 0$ ($T_c$), the onset of SC transition ($T_{c}^{on}$), defined at the point where $R_{sq}/R_N =0.95$, and the mean free path ($l$), estimated based on the $R_N$ value. For polycrystalline films we use $\rho_N l = const$, where $\rho_N = R_N d$ is the resistivity, and $const = 3.7\times10^{-12} \Omega$ cm$^2$ for crystalline niobium\cite{Mayadas1972}. For amorphous films a modification relating $const$ to residual resistivity has been proposed \cite{Gurvitch1986}, and we use it for our estimate. Despite this modification the $l$ values in the thinnest films fall below the average atomic distances, indicating that $l$ is underestimated, probably due to differences between actual band structure parameters and those used in Ref.\cite{Gurvitch1986}. The remaining parameters listed in the table are: the SMT critical field $B_c$ (at the lowest $T$), upper critical field $\mu_0 H_{c2}(0)$, and coherence length $\xi(0)$; they are estimated as explained further in the text.

\begin{figure}[t]
\centering
\includegraphics[width=0.9\textwidth]{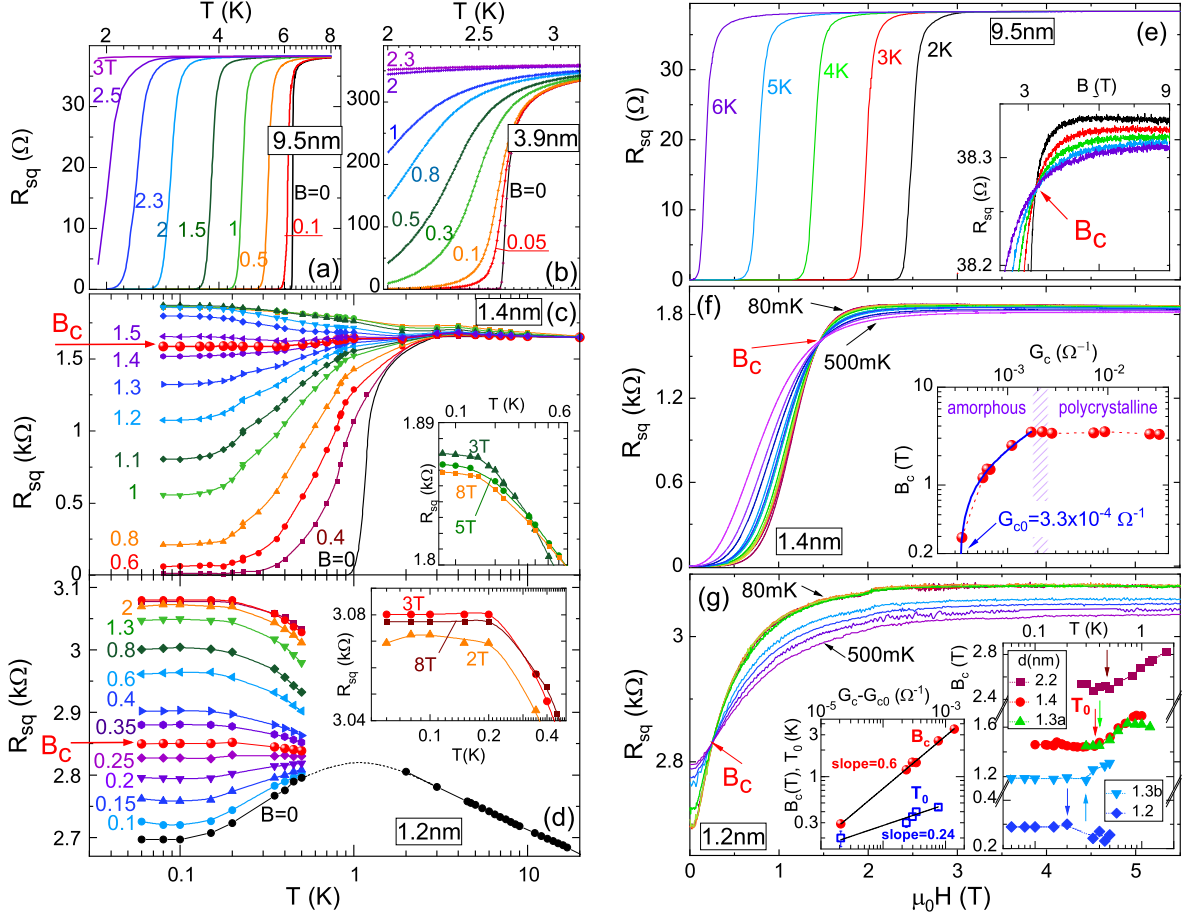}
\caption {$R_{sq}$ vs $T$ for various $B$ as labeled near the curves (in teslas) (a-d), and $R_{sq}$ vs $B$ at various $T$ as labeled near the curves (e-g), for films with various $d$: (a) 9.5 nm, (b) 3.9 nm, (c) 1.4 nm, (d) 1.2 nm, (e) 9.5 nm, (f) 1.4 nm, (g) 1.2 nm.
Red arrows indicate $B_c$. Insets in (c) and (d): regions of negative MR on the expanded scale. Inset in (e): expanded region near $B_c$. Inset in (f): $B_c$ vs $G_c$ (measured at the lowest $T$). Blue curve shows the fit described in the text, $B_c = A (G_c - G_{c0})^p$. Right inset in (g): $B_c$ vs $T$ for several amorphous films. The arrows indicate $T_0$, below which $B_c$ is constant. Left inset in (g): $B_c$ and $T_0$ vs $G_c - G_{c0}$ on a double logarithmic scale. The lines have slopes of 0.6 and 0.24, respectively.}\label{Rsq}
\end{figure}

Figs.\ref{Rsq}(a)-(d) show the dependencies of $R_{sq}(T)$ for various $B$ for films with $d$ equal to 9.5 nm (a), 3.9 nm (b), 1.4 nm (c) and 1.2 nm (d). The SC transition at $B = 0$ is present in all films with $d \geq 1.3$ nm, with the width increasing with decreasing $d$ from 0.1 K up to about 2 K. Fig.\ref{Rsq}(d) shows that even in the film with $d = 1.2$ nm non-complete SC transition is seen below 1 K. With the increase of the magnetic field the SC transition broadens, and the broadening grows with the decrease of $d$. In polycrystalline samples below the $T_c$ the resistance shows activated behavior, with activation energies consistent with collective vortex pinning (see Fig.S1 and discussion in Supplementary information). On the other hand, on cooling of the amorphous films, after the initial activated region, the resistance saturates at the finite, $B$-dependent level, which is smaller than the normal-state resistance\cite{sat} [Fig.\ref{Rsq}(c)]. Similar saturation has been observed for many other films with the SMT \cite{Ephron1996,Mason1999,Chervenak2000,Qin2006,Seo2006,Lin2012,Liu2013,Saito2015,Tsen2016,Tamir2019}. The origins of the resistance saturation is a hotly debated issue. While in case of some materials it has been attributed to marginal stability of the SC state against external noise \cite{Tamir2019}, the other possible explanations include formation of anomalous metallic phase, whose nature is still under debate \cite{Shimshoni1998,Das2001,Phillips2003,Galitski2005,Spivak2008,Kapitulnik2019}. The origins of saturation in the present case will be determined by future studies.

The increase of the magnetic field leads eventually to a weakly disordered metal, with negative $dR_{sq} /dT$ at low temperatures [Fig.\ref{Rsq}(c)]. However, when $B$ exceeds about 5 T, negative MR appears at the lowest $T$, as it is clearly seen in the insets to Figs.\ref{Rsq}(c) and \ref{Rsq}(d). Such negative MR at high magnetic field has been reported for weakly disordered films $a$-Mo$_x$Si$_{1-x}$ \cite{Okuma1998} and Nd$_{2-x}$Ce$_x$CuO$_{4+y}$ \cite{Gantmakher2003}, and attributed to the presence of the FCP on the nonsuperconducting side of the SMT \cite{Gantmakher2010}.

The characteristic feature of SMT transition is that at some critical magnetic field, $B_c$, and critical resistance, $R_c$, the derivative $dR_{sq} /dT$ switches from being positive to negative. The $B_c$ is indicated by red arrows in Figs.\ref{Rsq}(c) and \ref{Rsq}(d). This type of behavior produces crossing of the isotherms on the $R_{sq}(B)$ graphs. The examples are presented in Figs.\ref{Rsq}(e)-(g) for films with $d$ equal to 9.5 nm (e), 1.4 nm (f), and 1.2 nm (g). We observe such crossings in all Nb films with $d \leq 11.3$ nm. The $B_c$ (in the limit of lowest measured $T$) is almost $d$-independent in all polycrystalline films, equal to about 3.4 T, while it is rapidly reduced in amorphous films with decreasing film thickness. In contrast, the $R_c$ increases monotonously with the decrease of $d$ in all films, reaching the value of about 2.85 k$\Omega$ in the film with $d = 1.2$ nm. This value is much smaller than the quantum resistance for Cooper pairs, $R_Q \equiv h/4e^2 \approx 6.45$ k$\Omega$, at which the SIT is predicted to occur in the bosonic scenario \cite{Fisher1990}.

The relation between $B_c$ and $R_c$ for all films (at lowest $T$ measured in each film) may be summarized on a log-log plot of the $B_c$ versus critical conductance, $G_c = 1/R_c$ [inset to Fig.\ref{Rsq}(f)]. Qualitatively different behavior is seen in the polycrystalline and in the amorphous films. While in the polycrystalline films the $B_c$ is very weakly dependent on $G_c$, in the amorphous region it is well described by the power law shown by blue line, $B_c = A (G_c - G_{c0})^p$, where $A$ is constant and $p = 0.6 \pm 0.03$. The value of $G_{c0} = 3.3 \times 10^{-4}$ ${\Omega}^{-1}$, at which the $B_c$ reaches zero, is slightly smaller than the conductance of the $d = 1.2$ nm film (equal to 3.5 $\times 10^{-4}$ ${\Omega}^{-1}$). Therefore, $G_{c0}$ most likely has a meaning of the conductance of a metallic, non-SC background. The power-law dependence of $B_c$ on the film conductance is predicted by a model of disordered array of SC puddles coupled by a metallic background through a proximity effect \cite{Spivak2008}. Since in our films with decreasing $d$ a growing contribution of electron carriers to the conductance is detected at low $T$ \cite{Zaytseva2014}, it is possible that these carriers form metallic, non-SC background. We note, however, that the exponent of power law $p$ observed here differs from exponents calculated in Ref.\cite{Spivak2008}, which are predicted to be 1 or 1/4 for low and high film conductance, respectively.

Interestingly, in all amorphous films the $B_c$ exhibits peculiar behavior displayed in the right inset to Fig.\ref{Rsq}(g). Namely, the crossings of the consecutive pair of isotherms, which define the $B_c$, remain constant within experimental accuracy below some temperature $T_0$, which decreases with decreasing $d$. However, at higher temperatures the $B_c$ shifts to higher value for most of the films, except for the film with $d = 1.2$ nm (see Supplementary information for a detail example). The left inset shows that the $T_0$ depends as a power-law on the film conductance, what mimics power-law dependence of $B_c$ on $G_c - G_{c0}$, but with the exponent which is about twice as small. The shift of the $B_c$ has been reported for some systems, either as a result of sequential SIT's \cite{Biscaras2013,Shi2014,Zhang2018}, or due to Griffiths singularity \cite{Xing2015,Shen2016,Xing2017,Saito2018,Liu2019}. However, in most of these cases the $B_c$ is seen to shift to higher values on cooling, with the one exception of the reverse trend, similar to what we observe, reported recently for ultrathin crystalline lead films with reentrant resistance at low $T$ \cite{Liu2019}. The reentrant resistance is usually attributed to the presence of the FCP on the non-superconducting side of the SIT \cite{Gantmakher2003,Gantmakher2010,Parker2006}. What distinguishes these systems from our thin Nb films it that we do not observe reentrant resistance; instead, at the lowest $T$ ($T < T_0$) the resistance reaches a constant value $R_c$ at the SMT. Nevertheless, the peculiar upward shift of the $B_c$ with increasing $T$ suggests that SC correlations play important role in this behavior, and we will discuss this in detail below.

\subsection*{Upper critical field}

\begin{figure}[h]
\centering
\includegraphics[width=0.9\textwidth]{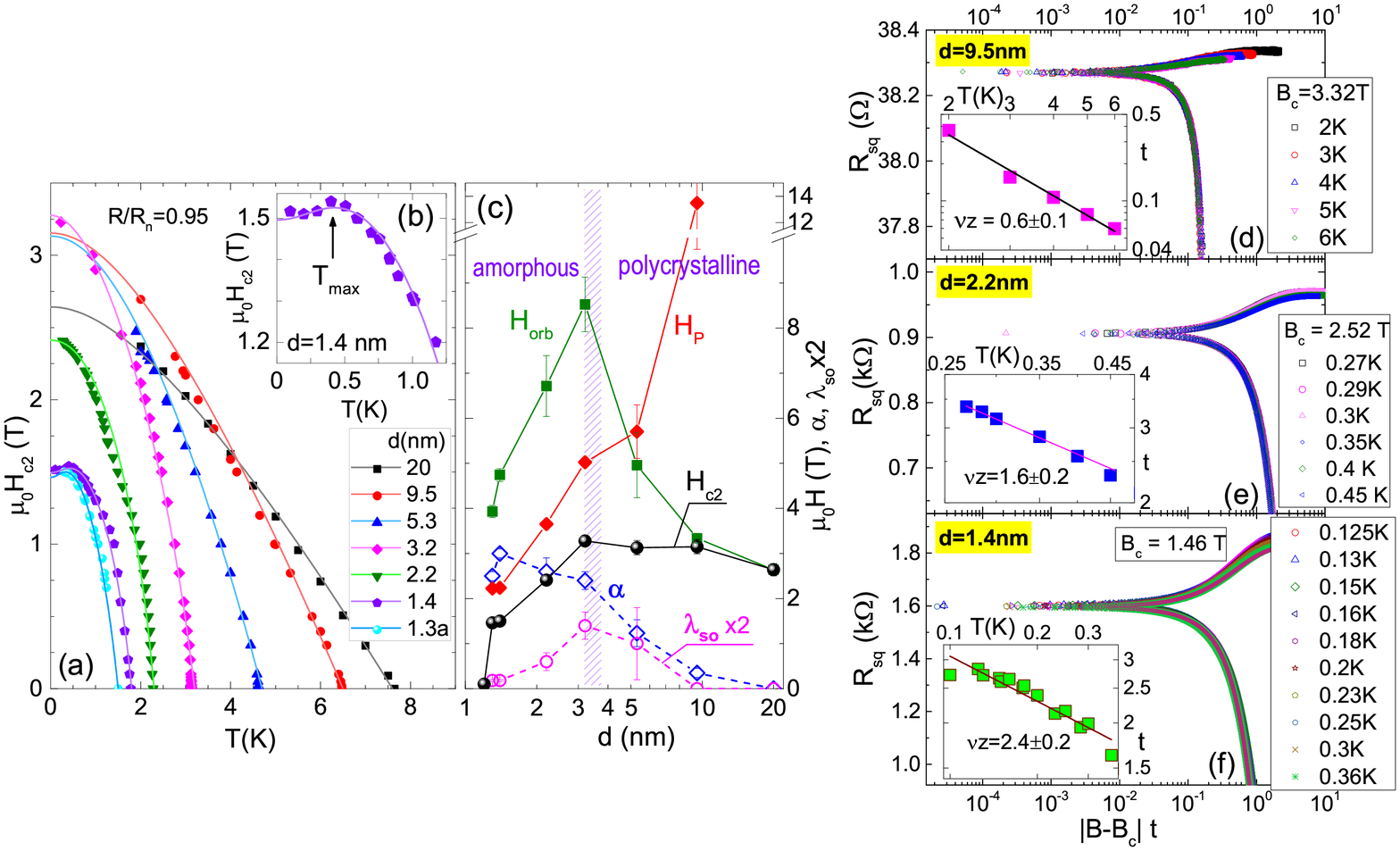}
\caption {(a) $\mu_0 H_{c2}$ vs $T$ for films with various $d$, as labeled in the figure. Solid lines show the fits to WHH theory. (b) The expanded low-$T$ portion of the data for $d = 1.4$ nm film. (c) The $d$-dependence of the fields: $H_{c2} (0)$ (spheres), $H_{orb} (0)$ (green squares), and $H_P (0)$ (red diamonds); and of Maki parameters $\alpha$ (open blue diamonds) and $\lambda_{so}$ (open magenta circles). (d-f) $R_{sq}$ as a function of scaling variable, $|{B - B_c}|t$, for films with $d$ = 9.5 nm (d), $d$ = 2.2 nm (e), $d$ = 1.4 nm (f). The insets show $t$ vs $T$ on a log-log scale. }\label{Hc2scal}
\end{figure}

It is important at this point to estimate the upper critical field, $H_{c2}$. We define $H_{c2}$ using the point of the SC onset ($R_{sq}/R_N = 0.95$). This eliminates the influence on the $H_{c2}$ of the vortex-related broadening of the transition, which is very large here. Fig.\ref{Hc2scal}(a) shows $H_{c2} (T)$ for representative films. We observe that the slope of the $H_{c2} (T)$-line at $T_c$, $(dH_{c2}/dT)_{T_c}$, increases with decreasing $d$ in polycrystalline films, and slightly decreases in amorphous films. Since the slope is proportional to $N_F \rho_N$, where $N_F$ is the density of states \cite{Gurevich2003}, the initial increase is most likely caused by the increase of $\rho_N$ due to enhanced surface scattering, while the slight decrease in amorphous films must be related, in addition, to the decrease of $N_F$, which compensates the increase of $\rho_N$. We also see that $H_{c2} (0)$ is rapidly reduced in the amorphous films, and in the thinnest films at the lowest $T$ a small region of double-valued $H_{c2}$ appears, as shown in detail for film with $d = 1.4$ nm in Fig.\ref{Hc2scal}(b), where $H_{c2}$ displays maximum at $T_{max} \simeq 0.42$ K.

Solid lines show the fits to the data using the conventional one-band, dirty limit WHH theory \cite{WHH1966}, which takes into account both spin paramagnetism and spin-orbit scattering through the Maki parameters $\alpha$ and $\lambda_{so}$ \cite{Maki1966}. In the absence of spin-paramagnetic effect and spin-orbit interaction ($\alpha=0$ and $\lambda_{so}=0$) the orbital pair breaking limits the upper critical field, which is given by $H_{orb}(0) \equiv H_{c2}(0)= -A(dH_{c2}/dT)_{T_c}T_c$, with the prefactor $A$ proportional to $2\Delta/kT_c$. In the presence of spin-paramagnetic effect the upper critical field is modified according to $H_{c2}(0) = H_{orb}(0)/\sqrt{1+\alpha^2}$, where $\alpha = \sqrt{2} H_{orb}(0)/H_P (0)$, and $H_P (0)$ is the zero-temperature paramagnetically limited field \cite{Maki1966}. In our calculation $A$ is adjusted to account for an increase of $2\Delta/kT_c$ with decreasing Nb film thickness (from 3.9 in polycrystalline films to 4.5 in ultrathin amorphous films) \cite{Park1986}, while $\alpha$ and $\lambda_{so}$ are treated as adjustable parameters. Thanks to wide $T$-range of the data in amorphous films, the Maki parameters may be estimated with reasonable accuracy\cite{note} (see WHH formulas and Table S1 in Supplementary information).

Based on the fits, we estimate $H_{c2}(0)$, $H_{orb}(0)$, and $H_P (0)$, which are plotted as a function of $d$ in Fig.\ref{Hc2scal}(c), together with Maki parameters. The zero-temperature coherence length $\xi(0)$, calculated using Ginsburg-Landau formula $\mu_0 H_{c2}(0)=\Phi_0/2\pi\xi^2(0)$ with $\Phi_0=2.07\times 10^{-15}$Wb, is listed in Table \ref{tab}. The $\xi(0)$ is larger than $d$ for all films with $d \leq 9.5$ nm, confirming that the films are 2D. The plots indicate the initial increase of $H_{c2} (0)$ with the decreasing $d$ in polycrystalline films, followed by a rapid decrease in amorphous films. This dependence is related to the interplay of $H_{orb}$ and $H_P$. While the $H_P (0)$ is decreasing monotonously with decreasing $T_c$, $H_{orb} (0)$ increases sharply on the approach to polycrystalline/amorphous boundary due to enhanced surface scattering. As a result, while in thicker polycrystalline films the orbital pair breaking dominates, on the approach to amorphous region the $H_{c2}$ becomes paramagnetically limited.

We note also that the dependence of $\lambda_{so}$ on $d$ follows essentially the dependence of $H_{orb}$ on $d$ (which is reasonable), that is, $\lambda_{so}$ is negligible for the thickest films, becomes the largest at the polycrystalline-amorphous boundary, and it is strongly suppressed in the thinnest films. According to WHH theory \cite{WHH1966} the $H_{c2}$ should become double-valued when Maki parameter $\alpha$ exceeds the value $\alpha_c = (1+1.589{\lambda_{so}}/{\lambda^c})/(1-{\lambda_{so}}/{\lambda^c}$), where $\lambda^c = 0.539$. This condition is fulfilled in the present experiment for the thinnest films, in which we indeed observe double-valued $H_{c2}$.

\subsection*{Scaling analysis}

We now turn attention to the scaling analysis. We use conventional assumption that the magnetic field $B_c$, which is the crossing point of the isotherms, defines the critical field, at which quantum phase transition (QPT) may occur in the $T=0$ limit. It is predicted that in the vicinity of the QPT at $T = 0$ the spatial correlation length $\xi$ and the dynamical correlation length $\xi_{\tau}$ diverge as a power law on the approach to the critical point $B_c$, $\xi \propto |{\delta}|^{-\nu}$ and $\xi_{\tau} \propto \xi^z$, where $\delta = B - B_c$, and $\nu$ ($z$) is spatial (dynamical) critical exponent \cite{Fisher1990,Sondhi1997}. At $T \neq 0$ the time dimension is limited by temperature fluctuations, what introduces $T$-dependent dephasing length, $L_\phi \propto T^{-1/z}$, beyond which quantum fluctuations lose phase coherence. This leads to a prediction that in the critical region all relevant quantities are universal functions of the scaling variable $|{\delta}|T^{-1/{\nu}z}$. Here, in order to verify scaling hypothesis, we use two methods \cite{Sondhi1997,Hebard1990}.

(1) We test if the resistance data, measured at various fixed temperatures in the vicinity of $B_c$ may be collapsed on a single curve given by $R_{sq}(B,T) = R_c f(|{B - B_c}|t)$ by adjusting the parameter $t(T)$ at each temperature, where $t$ should follow the power law, $t(T) = T^{-1/{\nu}z}$. Figs.\ref{Hc2scal}(d)-(f) show examples of this analysis for films with $d = 9.5$ nm, 2.2 nm, and 1.4 nm.

(2) We calculate the partial derivative, $(\partial R_{sq}/\partial B)_{B_c} \propto T^{-1/{\nu}z}$, and plot it versus $1/T$ to extract the slope, which gives inverse of the product of critical exponents ${\nu}z$. This is illustrated in Fig.S4 in Supplementary information.

As shown in Fig.\ref{Hc2scal}(d), in the case of polycrystalline film with $d = 9.5$ nm a good collapse of the data is obtained in a broad $T$-range, with a single critical field, and product of critical exponents ${\nu}z = 0.6 \pm 0.1$. This is confirmed by method (2), from which we obtain the same value of ${\nu}z$ for all polycrystalline films. However, this is not the case for amorphous films. As revealed in Figs.\ref{Hc2scal}(e-f), for these films good collapse of the data, with a single $B_c$ and $R_c$, is limited to low temperatures, $T < T_0$, where $T_0$ is the temperature below which the $B_c$ is constant. The values of critical exponents vary from film to film, but they are always larger than 1. Method (2) indicates that in case of amorphous films the partial derivative data follow two different slopes in different $T$-regions, low $T$ ($T<T_0$), and high-$T$ ($T>T_0$). In the low-$T$ region the average value of the critical exponent extracted from the slope is ${\nu}z = 2.2 \pm 0.2$ for five amorphous films with different $d$ (see Fig.S4 in Supplementary information). Curiously, the slope for amorphous films in the high-$T$ range is identical to the slope for polycrystalline films, leading to low value of "apparent critical exponent". However, the very fact that the $B_c$ and $R_c$ are changing in high-$T$ region suggests that at $T > T_0$ one-parameter scaling breaks down in amorphous films \cite{Gantmakher2010}. The possible origin of the curious "apparent critical exponent" is further discussed in next sections.

The small value of ${\nu}z$ observed in polycrystalline films is close to 2/3. Such ${\nu}z$ has been observed in conventional 2D superconducting films, for example, in a-Bi \cite{Markovic1998} or a-NbSi \cite{Aubin2006}. This is consistent with (2+1)D $XY$ model for a 2D superconductor in the clean limit, provided that $z = 1$, which is the usual assumption in the case of the system with long-range Coulomb interactions between charges \cite{Fisher1990,Sondhi1997}, confirmed experimentally in several materials \cite{Hebard1990,Yazdani1995}. The clean limit corresponds in this case to weakly disordered vortex matter, which in polycrystalline films at low temperatures and low magnetic fields creates pinned vortex glass phase. Upon increasing of the field it unpins at fields very close to $H_{c2}$, as discussed in Supplementary information. On the other hand, in the amorphous films at low $T$ the ${\nu}z$ exceeds 1. Such ${\nu}z$ is predicted by (2+1)D $XY$ model for 2D disordered systems \cite{Fisher1990,Sondhi1997}, and has been observed in a variety of disordered films \cite{Hebard1990,Seidler1992,Yazdani1995,Gantmakher2000,Baturina2004,Bielejec2002,Steiner2008}. In this case at low $T$ and in the weak magnetic field the disorder leads to the formation of strongly disordered vortex glass phase, which melts upon the increase of the magnetic field at fields substantially lower than $H_{c2}$.

\subsection*{Superconducting fluctuations above $H_{c2}$}

\begin{figure}[t]
\centering
\includegraphics[width=0.9\textwidth]{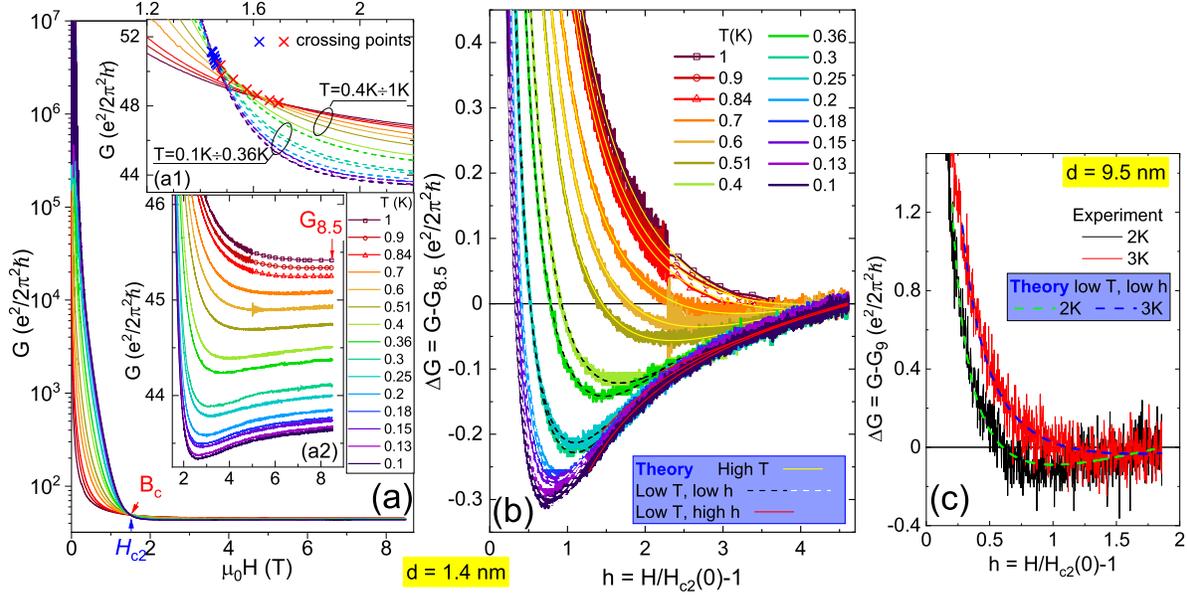}
\caption{(a) $G$ vs magnetic field for labeled temperatures in film with $d = 1.4$ nm. (a1) expanded area in the vicinity of $B_c$. Low-$T$ (high-$T$) data are indicated by dashed (continuous) lines and blue (red) crosses, respectively. (a2) region of high magnetic fields. (b) $\Delta G$ versus normalized magnetic field $h$ at labeled temperatures for $d = 1.4$ nm. The experimental data are shown by thick color lines. Thin lines show theoretical asymptotic dependencies from Ref.\cite{Glatz2011}, fitted in the region of high-$T$ (continuous yellow lines), region of low-$T$ and low $h$ (dashed lines, white or black), and region of low-$T$ and high $h$ (red lines - note that for clarity only some red lines are plotted). (c) $\Delta G$ versus $h$ at labeled temperatures for $d = 9.5$ nm ($G_9$ is the conductivity at 9 tesla). The dashed lines show the fitted expressions from low-$T$, low $h$ region.
} \label{Delta}
\end{figure}

In order to elucidate puzzling feature of small "apparent critical exponent" observed at high $T$ in the amorphous films, we have examined in more detail the influence of the magnetic field on the conductance $G = 1/R_{sq}$ of the film with $d=1.4$ nm, for temperatures below the $T_c$. The results are displayed in Fig.\ref{Delta}(a), where $G$ is shown in units of $G_0 = {e^2}/2{{\pi}^2}\hslash$. At zero magnetic field the $G$ is very large for all temperatures ($>10^7$), suggesting the approach to long-range SC order. Weak magnetic field ($H \ll H_{c2}$) suppresses superconductivity rapidly at high $T$, and less rapidly at low $T$. However, this trend is reversed at high field, in the vicinity of $H_{c2}$, as illustrated in Figs.\ref{Delta}(a1)-(a2). Fig.\ref{Delta}(a1) presents expanded area in the vicinity of crossing points ($B_c$). Dashed lines and blue crosses indicate low $T$ data ($0.1 \leq T \leq 0.36$ K), while continuous lines and red crosses indicate high $T$ data ($0.4 \leq T \leq 1$ K). We observe that in the low-$T$ region $G$ is very rapidly reduced with increasing field, what produces very small spread of blue crossing points. On the other hand, the suppression of $G$ by the magnetic field is much slower in the high-$T$ range, what results in large spread of red crossing points. Thus, the absence of well defined $B_c$ and $R_c$ at $T > T_0$ is traced to slow suppression of the conductance by the magnetic field at high $T$.

Fig.\ref{Delta}(a2) shows the expanded region of conductance in high magnetic fields ($H > H_{c2}$). At high $T$ (top curves) the $G$ is gradually reduced with increasing field, until it saturates at high-field value $G_{8.5}$. Such a behavior is expected if the magnetic field suppresses SC fluctuations, so that at high field the conductance of the normal state is reached. However, when $T$ is reduced, a minimum appears on $G(H)$ curve at intermediate field, followed by subsequent slow increase of $G$ up to $G_{8.5}$. With lowering of $T$ the minimum moves towards lower magnetic fields, and becomes more pronounced. The appearance of such minimum has been predicted by theories of the FCP above $H_{c2}$ in 2D superconductors at very low temperatures, in the region of quantum fluctuations \cite{Galitski2001,Glatz2011,Varlamov2018}. In order to compare our experimental results to theory, we replot the data in Fig.\ref{Delta}(b) as $\Delta G = G - G_{8.5}$ versus normalized magnetic field, $h = H/H_{c2}(0)-1$. We consider $G_{8.5}$ as beeing close to normal-state conductance, so that $\Delta G (h)$ is mostly due to superconducting fluctuations of the FCP.

The first analysis of quantum fluctuations at $H > H_{c2}$, in a lowest Landau level approximation (valid close to the $H_{c2} (T)$) is due to Galitski and Larkin (GL), who find negative contribution of SC fluctuations to conductivity in a narrow range close to zero temperature \cite{Galitski2001}. Subsequently, exact calculations in the first order perturbation theory by Glatz et al. (GV) \cite{Glatz2011,Varlamov2018} have provided formulas for the full $H-T$ phase diagram above the $H_{c2}$. Various contributions to fluctuation conductivity have been considered, i.e., Aslamazov-Larkin, Maki-Thompson (MT), single-particle density of states, and one-electron diffusion coefficient renormalization (DCR) terms. Full evaluation of GV formulas require rather tedious numerical calculations, which have been so far compared to experimental data in just a few cases \cite{Glatz2011,Varlamov2018}. Here, we have tested both GL and GV formulas for description of our data. The details of the fitting procedures are provided in Supplementary information, together with comparison of the GL theory to experimental data. In Figs.\ref{Delta} (b) and (c) we show data for two films, $d = 1.4$ nm and $d = 9.5$ nm, respectively, together with the fitted GV formulas (for simplicity we use asymptotic forms only). The basic conclusion is that negative contribution of SC fluctuations to conductivity above the $H_{c2}$ is well described by theory of the FCP. It is also significantly more pronounced in the case of amorphous film, particularly in the temperature range below $T_0$.

\subsection*{Phase diagrams}

\begin{figure}[h]
\centering
\includegraphics [width=0.95\textwidth]{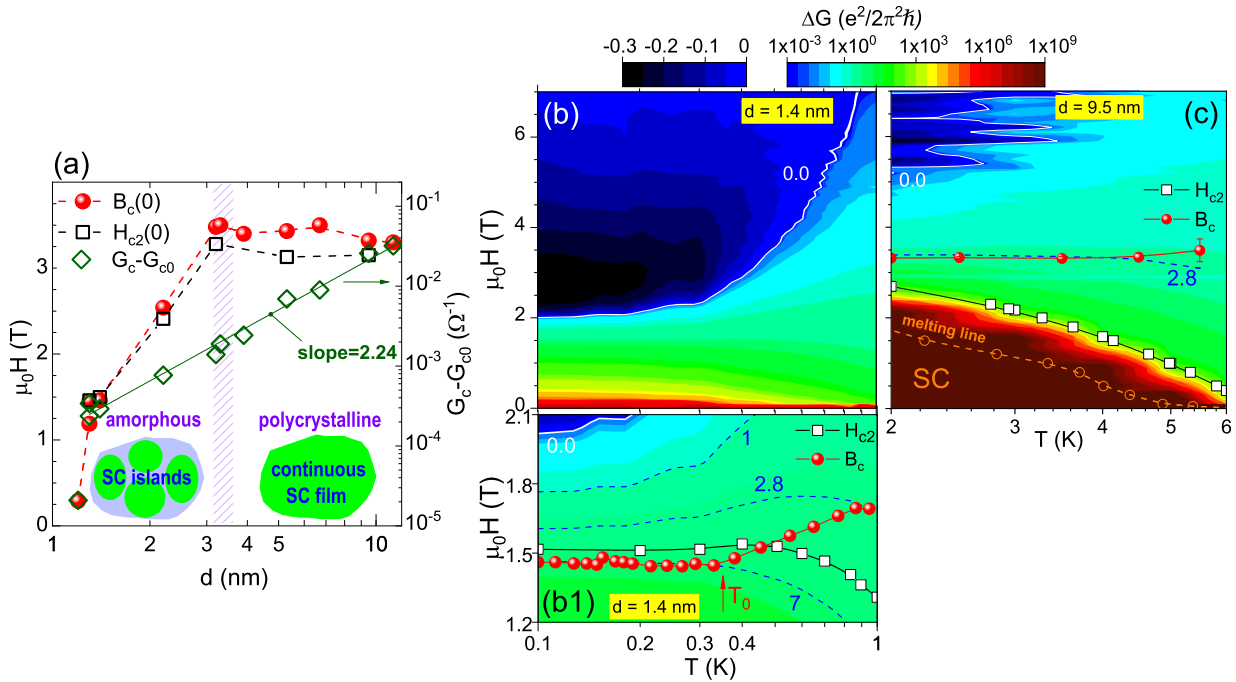}
\caption{(a) $T=0$ phase diagram in $B-d$ plane for ultrathin Nb films:  $B_c (0)$ (spheres) and $H_{c2} (0)$ (squares) (\textit{left scale}), and $G_c - G_{c0}$ (diamonds, \textit{right scale}). The insets show schematically the low-$T$ SC in the amorphous (left) and polycrystalline (right) films. (b-c) Color maps of $\Delta G (H,T)$ for films with $d=1.4$ nm (b) and $d=9.5$ nm (c); (b1) shows expanded portion of (b). The scale is logarithmic for positive $\Delta G$ (rainbow colors), and linear for negative $\Delta G$ (blue-black); dashed blue lines are drawn at labeled values of $\Delta G$. Points indicate $H_{c2}$ (white squares), $B_c$ (red spheres), and melting line (open orange circles and dashed orange line).}
\label{Sum}
\end{figure}

We now summarize our results by presenting phase diagrams. First, we construct a $T=0$ phase diagram of Nb films in the $B-d$ plane (Fig.\ref{Sum}(a), $d$ on a logarithmic scale). The left (linear) scale shows $H_{c2} (0)$ and the $B_c (0)$ (which is $B_c$ in the limit of the lowest $T$ measured in each film). Note that in polycrystalline films the $B_c (0)$ is somewhat higher than the $H_{c2}(0)$, what likely results from arbitrary definition of $H_{c2}(0)$ \cite{note}. However, the $B_c (0)$ shifts slightly below $H_{c2}(0)$ in thinnest amorphous films because of the shift of $B_c$ to lower values at $T < T_0$. On the right (logarithmic) scale we show film conductance $G_c - G_{c0}$. The continuous green line shows the fit to the data, which indicates that for $d \geq 1.3$ nm the conductance follows a power law dependence on the film thickness, $G_c - G_{c0} \sim d^{2.24}$, consistent with the surface scattering in thin films, as has been previously discussed \cite{Zaytseva2014}. $G_c - G_{c0}$ drops by an order of magnitude at $d = 1.2$ nm, in accordance with the fact that only non-complete SC transition survives in this film. As already mentioned, the $B_c ({G_c})$ dependence suggests a possibility that in the amorphous films at low $T$ inhomogeneities develop, in the form of disordered arrays of SC islands coupled via metallic background by proximity effect \cite{exponent}. On the other hand, no sign of such inhomogeneities exist in polycrystalline films. We sketch this behavior schematically in the insets of Fig.\ref{Sum}(a), where green are the SC regions, and light blue depicts the metallic, non-SC background.

Next, in Figs.\ref{Sum}(b-c) we present $H-T$ phase diagrams for two films, amorphous with $d = 1.4$ nm (b-b1), and polycrystalline with $d = 9.5$ nm (c). The background shows color maps of $\Delta G (H,T)$, with positive $\Delta G$ (rainbow colors) on logarithmic scale, and negative (blue-black) on linear scale. The dark red-to-brown region (with $\Delta G \gtrsim 1 \times 10^5$) at the bottom indicates the approach to SC zero-resistance state. In (c) this state is a pinned vortex glass, which covers large portion of the phase diagram. From the Arrhenius plots of the resistance we determine the melting line (as described in Supplementary information), shown here by open orange circles and dashed line. In (b), on the other hand, this region is barely seen at the lowest magnetic field, confirming that the zero-resistive state is limited to the lowest magnetic fields at best. However, the melting line may still be observed in the amorphous film with $d = 2.2$ nm (see comment in Supplementary Information).

The blue-to-black regions at the left top corner indicate the areas with negative contribution of SC fluctuations due to the FCP. In (c) this region is rather small, and restricted to high magnetic fields. In (b), on the other hand, the depression is much better defined, and influences conductivity in a very large portion of the phase diagram. Nevertheless, at high $T$ the depression in amorphous film disappears, the $\Delta G$ becomes positive, and extends to high magnetic fields, similar as in polycrystalline film. We believe that this feature explains the observation of small "apparent critical exponent" in amorphous films at high $T$, that is, simply the dependence of partial derivative $(\partial R_{sq}/\partial B)_{B_c}$ on $1/T$ becomes similar to that seen in polycrystalline films. Moreover, if we now plot the dependence $B_c (T)$ on the maps [red spheres in Figs.\ref{Sum}(b1-c)] we see that in polycrystalline film it is constant and located at $\Delta G \simeq 2.8$. Notably, the location of the $B_c$ in amorphous film is also at $\Delta G \simeq 2.8$ at high $T$. However, at low $T$ ($T < T_0$) the position of the $B_c$ is pushed to higher value of positive $\Delta G \simeq 7$ (lower magnetic field). This shift suggests that the nature of SC correlations in the amorphous film changes as the temperature is decreased towards $T_0$.

We propose to explain this evolution by a competition between proximity coupling in a disordered array of SC islands, which promotes long-range SC phase coherence, and low-$T$ quantum fluctuations, enhanced in the presence of magnetic field, which break phase coherence. Following experiments on arrays of mesoscopic SC islands on metallic underlayer \cite{Eley2012,Han2014} we assume two energy scales in the system, $J$, related to coupling between grains on an individual island, and $J'$, the coupling between neighboring islands, which takes standard proximity form, $J' \approx J'_0 \exp{(-t / \xi_N (T))}$. Here $J'_0$ is the amplitude, $t$ is the distance between islands, and $\xi_N (T) \sim \sqrt{\hslash D/(k_B T)}$ is the normal-metal coherence length with diffusion constant $D$. Depending on the geometry of the SC array and the normal metal parameters, this form of $J'$ leads either to gradual enhancement of coupling between islands on decreasing $T$, what results in two-step SC transition \cite{Eley2012}, or to the decoupling of SC islands due to quantum fluctuations at low $T$, what produces metallic $T = 0$ state \cite{Han2014}. In the presence of perpendicular magnetic field proximity couplings $J'$ are suppressed exponentially beyond the magnetic length $L_H = \sqrt{{\Phi}_0/H}$ (${\Phi}_0$ is a flux quantum), what additionally enhances quantum fluctuations, and promotes the decoupling of SC islands.

Applying this model to our system, we presume that polycrystalline films consist of one large island, with only $J$ coupling between grains contributing to one-step transition (which is broadened due to surface-related disorder and vortex activation). On the other hand, in amorphous films disordered arrays of SC islands develop, with electron carriers detected at low $T$ by Hall effect \cite{Zaytseva2014} playing the role of metallic layer. The distribution of array parameters within film, and, therefore, distributions of $J'$ couplings, smears SC transitions. Nevertheless, we expect to observe higher-$T$ region in the vicinity of the $T_c$ onset, in which $J$ leads to coupling within each grain, and, on decrease of $T$, the approach to low-$T$ region with increasingly relevant $J'$ couplings. On the decrease of film thickness $d$, the average $t$ increases, and $\xi_N$ decreases, what leads to gradual decrease of the average $J'$. In the thinnest films this produces barely visible zero-resistive state in the absence of magnetic field [Fig.\ref{Sum}(b)], while in the presence of the field $J'$ couplings are suppressed, and the enhancement of quantum fluctuations on lowering of temperature results in gradual suppression of phase coherence between islands, and, therefore, gradual decrease of the $B_c$  [Fig.\ref{Sum}(b1)]. We believe that this scenario explains the observation of two quite distinct regions of SC correlations visualized in Figs.\ref{Sum}(b-b1). Finally, we note that the enhancement of quantum fluctuations may be responsible for the saturation of resistance, which we observe in the thinnest films. However, we expect to verify this issue by further studies.

\section*{Conclusion}

The $H_{c2}$ and the $B$-induced SMT have been examined in ultrathin Nb films, which undergo a transition from polycrystalline to amorphous structure on the decrease of thickness. The properties of the SC state in the amorphous films are very different from those in polycrystalline films. The $H_{c2}$ is found orbitally limited in the first case, and paramagnetically limited in the latter. The scaling analysis indicates distinct scaling exponents in these two types of films, consistent with (2+1)D \textit{XY} model for 2D superconductor in the clean or dirty limit, respectively. The negative contribution of SF fluctuations to conductivity is found above $H_{c2}$, well described by theories of fluctuating Cooper pairs; this contribution is much more pronounced in the amorphous films. All these differences strongly suggest that local inhomogeneities develop in amorphous films, in the form of proximity-coupled SC islands. On the decrease of temperature SC correlations in amorphous films evolve in an unusual fashion, suggesting the suppression of proximity-induced phase coherence by quantum fluctuations.

\section*{Methods}

The $R_{sq}$ as a function of temperature and magnetic field up to 9 T, perpendicular to the film plane, was measured on a lithographically patterned resistance bridge using a standard four-probe method, with dc current for $T > 2$ K and low-frequency ($f = 19$ Hz), $ac$ lock-in techniques with $I = 10$ nA for lower $T$, down to 80-100 mK. Care was taken to remain in the Ohmic regime. To cover various $T$-ranges in different samples several different cryostats were used, He-4 or PPMS for high$-T$ range, $^3$He cryostat for 0.3 K $ < T < 2$ K, and LHe dilution refrigerator with low-pass filters for $T < 1$ K. In the mK-range the data were accumulated during the field sweep, while at higher $T$ both the field, and the temperature sweeps were used. More details are available in Supplementary information.

\section*{Acknowledgements}

We are grateful to Leyi Y. Zhu and Chia-Ling Chien (Johns Hopkins University) for growing the films, and to G. Grabecki and J. Wr\'{o}bel (Institute of Physics, PAS) for providing experimental apparatus for mK measurements. The work has been supported by Polish NSC Grants No. 2011/01/B/ST3/00462 and 2014/15/B/ST3/03889. The research was partially performed in the laboratory co-financed by the ERDF Project NanoFun POIG.02.02.00-00-025/09. The work at JHU has been supported by NSF grant DMREF1729555. B.C.C. acknowledges the National Science Center, Poland, Project. no. 2016/23/P/ST3/03514 within the POLONEZ programme. The POLONEZ programme has recieved funding from the European Union's Horizon 2020 research and innovation programme under the Marie Sklodowska-Curie grant agreement No. 665778.

\section*{Author contributions statement}

M.Z.C. and I.Z. conceived the experiments. I.Z., A.A. and Y.S. conducted the experiments. M.Z.C. and I.Z. analysed the results. B.C.C. performed part of numerical calculations. M.Z.C. wrote the manuscript with contributions from I.Z. All authors approved the manuscript before submission.

\section*{Competing interests}

The authors declare no competing interests.

%
%


\begin{thebibliography}{99}
\bibitem{note}
The definition of $H_{c2}$ at $R_{sq} /R_N = 0.95$ is arbitrary, and changing it slightly (for example, from 0.95 into 0.9) affects slightly the Maki parameters. However, the general behavior of parameters with reduction of $d$ is not affected.
\end{thebibliography}
\end{document}